\documentstyle[12pt,moriond,epsf,psfig,colordvi]{article}

\def\be{\begin{equation}}
\def\ee{\end{equation}}
\def\bea{\begin{eqnarray}}
\def\eea{\end{eqnarray}}
\def\beas{\begin{eqnarray*}}
\def\eeas{\end{eqnarray*}}

\newcommand{\nub}{\overline{\nu}}

\def\nim#1#2#3  {{\em Nucl. Instr. Meth.} {\bf#1}, #2 (#3) }
\def\np#1#2#3   {{\em Nucl. Phys.} {\bf#1}, #2 (#3) }
\def\pcps#1#2#3 {{\em Proc. Cam. Phil. Soc.} {\bf#1}, #2 (#3) }
\def\pl#1#2#3   {{\em Phys. Lett.} {\bf#1}, #2 (#3) }
\def\prep#1#2#3 {{\em Phys. Rep.} {\bf#1}, #2 (#3) }
\def\prev#1#2#3 {{\em Phys. Rev.} {\bf#1}, #2 (#3) }
\def\prl#1#2#3  {{\em Phys. Rev. Lett.} {\bf#1}, #2 (#3) }
\def\prs#1#2#3  {{\em Proc. Roy. Soc.} {\bf#1}, #2 (#3) }
\def\ptp#1#2#3  {{\em Prog. Th. Phys.} {\bf#1}, #2 (#3) }
\def\rmp#1#2#3  {{\em Rev. Mod. Phys.} {\bf#1}, #2 (#3) }
\def\rpp#1#2#3  {{\em Rep. Prog. Phys.} {\bf#1}, #2 (#3) }
\def\zp#1#2#3   {{\em Zeit. Phys.} {\bf#1}, #2 (#3) }
\def\epj#1#2#3   {{\em Eur. Phys. Jour.} {\bf#1}, #2 (#3) }
\begin{document}
\input psfig.tex
\newcommand{\linespace}[1]{\protect\renewcommand{\baselinestretch}{#1}
  \footnotesize\normalsize}
\begin{flushright} 
\typeout{need preprint number} 
UR-1534\\
ER-40685-918\\
Jul. 22, 1998
\end{flushright}

\title{ MEASUREMENTS OF THE LONGITUDINAL
STRUCTURE FUNCTION AND $|V_{cs}|$ IN THE CCFR EXPERIMENT
 \footnote{ To be published in proceedings of the 6th International
Workshop on Deep Inelastic Scattering and QCD, Brussels, Apr. 1998.}}

\author{
        U.~K.~Yang$^f$, C.~McNulty$^b$,
	C.~G.~Arroyo$^b$, L.~de~Barbaro$^e$, P.~de~Barbaro$^f$, 
	A.~O.~Bazarko$^b$, R.~H.~Bernstein$^c$, A.~Bodek$^f$, 
	T.~Bolton$^d$, H.~Budd$^f$, J.~Conrad$^b$, D.~A.~Harris$^f$, 
	R.~A.~Johnson$^a$, J.~H.~Kim$^b$, B.~J.~King$^b$, T.~Kinnel$^g$, 
	M.~J.~Lamm$^c$, W.~C.~Lefmann$^b$, W.~Marsh$^c$, 
	K.~S.~McFarland$^c$, S.~R.~Mishra$^b$, 
	D.~Naples$^d$, P.~Z.~Quintas$^b$, A.~Romosan$^b$, 
	W.~K.~Sakumoto$^f$, H.~Schellman$^e$, F.~J.~Sciulli$^b$,
	W.~G.~Seligman$^b$, M.~H.~Shaevitz$^b$, W.~H.~Smith$^g$,
	P.~Spentzouris$^b$, E.~G.~Stern$^b$, M.~Vakili$^a$,
	 J.~Yu$^c$ }

\vspace{0.15in}
\address{$^a$~University of Cincinnati, 
$^b$~Columbia University, \\ 
$^c$~Fermi National Accelerator Laboratory,
$^d$~Kansas State University, \\
$^e$~Northwestern University,
$^f$~University of Rochester,\\
$^g$~University of Wisconsin}


\begin{center} 
\end{center} 

\maketitle\abstracts{
Measurements of charged current neutrino and anti-neutrino nucleon 
interactions in the CCFR detector are used to extract the structure 
functions, 
$F_2$, $xF_3^\nu$, $xF_3^{\overline \nu}$ and $R$(longitudinal) 
in the kinematic region
$0.01<x<0.6$ and $1<Q^2<300$ GeV$^2$. The new measurements 
of  $R$ in the $x<0.1$ region provide a constraint on the level
of the gluon distribution.
The $x$ and $Q^2$ dependence of $R$ is compared with a QCD based fit
to previous data. The CKM matrix element $|V_{cs}|$ is extracted  
from a combined analysis of $xF_3$ and dimuon data.}

\section {Introduction}

Neutrino-nucleon scattering data is unique in probing the
valence quark, sea quark, and gluon content of the nucleon 
through measurements
of four different structure functions ( $F_2$, $xF_3^\nu$, 
$xF_3^{\nub}$ and $R$ ). In charged current
$\nu$-nucleon interaction, the differential cross section
can be written as follows:
$$
 \frac{d^2\sigma^{\nu (\nub)}}{Edxdy} \propto
   \left [ F_2^{\nu (\nub)}(x,Q^2)
   \left ( 1-y+ \frac {y^2}{2(1+R)} \right)
   \pm xF_3^{\nu (\nub)}(x,Q^2) \left( y-\frac{y^2}{2} \right) \right]
$$
where $x$ is the Bjorken scaling variable,
$Q^2$ is the square of the four momentum transfer, and $y$ is the fractional
lepton energy loss.
Within the quark-parton model (in leading order), 
$F_2^{\nu,\nub}=x\sum (q+\overline{q})$
and $xF_3^{\nu,\nub}=x\sum (q-\overline{q})^{+2x(s-c)}_{-2x(s-c)}$.
Thus, the measurements of $F_2$ and $xF_3$ provide
information on the valence and sea quarks.
The size of the strange sea at low $x$ can be obtained from 
the difference in $xF_3^{\nu}$ and $xF_3^{\nub}$.
The ratio $R$ of the longitudinal and transverse structure functions
provides information about the transverse momentum 
of the nucleon constituents. 
In leading order, $R = 0$, since the
quarks have no transverse momentum. In the next to leading order
formalism (NLO), $R$ is non-zero
because of transverse momentum associated with gluon emission~\cite{qcd1}.
The NLO QCD prediction is given 
by an integral over the quark and gluon distribution and
is proportional to $\alpha_s$.
Thus, a measurement of $R$ provides a test of
perturbative QCD at large $x$, and probes the gluon density 
at small $x$ where the quark contribution is small. 
However, previous measurements of $R$ have not been performed 
over wide kinematic regions
except in the high $x$ and low $Q^2$ regions where non-perturbative 
contributions are significant. A good measurement of $R$
requires high statistics
data at different beam energies (or $y$) for each 
$x$ and $Q^2$ bin.
Poor knowledge of $R$ , especially at small $x$,
leads to uncertainties in the extracted values
of the structure function, $F_2$.
Therefore, measurements of $R$ both at large $Q^2$ and at small $x$
are needed. The CCFR experiment has a unique
capability to measure $R$ in these kinematic
regions by using a high-intensity high energy wide-band neutrino beam
and the massive CCFR neutrino detector.

We report here on
new extractions of $R$ and the magnitude of strange sea
using a sample of neutrino and antineutrino
interactions in the CCFR  detector. An extraction of the CKM matrix
element $|V_{cs}|$ is achieved by combining the strange sea measurement
with our previous dimuon results~\cite{dimu}.

\section {A Preliminary CCFR measurement of $R$ and $|V_{cs}|$}

The event sample combines data
from two runs (E744 and E770) collected using
the Fermilab Tevatron Quad-Triplet neutrino beam. The wide-band beam
is composed of $\nu_\mu$ and $\nub_\mu$ with energies up to 600 GeV.
The CCFR neutrino detector~\cite{calib} consists of non-magnetic
steel-scintillator target calorimeter
instrumented with drift chambers.
The hadron energy resolution is 
$\Delta E/E = 0.85/\sqrt{E}$(GeV). The target is followed by a solid
iron toroid muon spectrometer which measures muon momenta with a resolution
$\Delta p/p = 0.11$. The
three independent kinematic variables $x$, $Q^2$, and $y$ are
reconstructed from measurements of
the hadronic energy ($E_h$), muon
momentum ($P_\mu$), and muon angle ($\theta_\mu$).
The relative flux~\cite{SEL}
at different energies is obtained from the events 
with low hadron energy, $E_h < 20$ GeV, and is normalized to
the average value of the 
world's neutrino total cross section measurements,
 $\sigma^{\nu N}/E=
(0.677\pm0.014)\times10^{-38}$ cm$^2$/GeV, and $\sigma^{\overline{\nu} N}
/\sigma^{\nu N}=0.499\pm0.005$~\cite{SEL}. The total data sample used 
in structure function extraction consists of
 1,280,000 $\nu_{\mu}$ and 270,000 $\nub_{\mu}$
 events (after fiducial and kinematic cuts $P_{\mu}>15$ GeV, 
$\theta_{\mu} <.150$,
$E_h > 10$ GeV, $Q^2 > 1$ GeV$^2$, and $30<E_{\nu}<360$ GeV). Dimuon
events are removed 
and used in an separate analysis
leading to another determination of the strange sea~\cite{dimu}.

The structure functions (SF's) are
extracted by minimizing a $\chi^2$ 
in the comparison of the $y$ distribution 
for data
and Monte Carlo (MC) events. In order to extract the SF's as a function of
$x$ and $Q^2$,  a $\chi^2$ analysis
is formed in each ($x$, $Q^2$) bin:
$$
\chi ^2(x,Q^2)=\sum_{20~y-bins}^\nu \frac{\left[ N_{data}^\nu-N_{MC}^\nu%
\left(SF \right)
\right]^2}{\sigma_{data}^2+\sigma_{MC}^2}+%
\sum_{20~y-bins}^{\overline{\nu }}\frac{%
\left[ N_{data}^{\overline{\nu }}-N_{MC}^{\overline{\nu }}
\left(SF \right)
\right] ^2}{\sigma_{data}^2+\sigma_{MC}^2}
$$
The effects of detector
acceptance and resolution smearing are implemented in the generation of
the MC events ($N_{MC}$).
The analysis incorporates
bin centering, normalization, and various corrections 
(e.g. QED and electroweak radiative corrections,
the $W$ boson propagator, the non-isoscalar nature of the iron target,
and the effects of the charm quark mass using the slow rescaling
formalism).

We obtain the values of $R$ and $\Delta xF_3$ ($=xF_3^\nu - xF_3^{\nub} $)
as well as $F_2$ and $xF_3^{avg}$ ($=(xF_3^\nu + xF_3^{\nub} )/2$)
using the above global fits ( after few iterations, 
because a priori knowledge of the SF values
is required to generate the MC events). The sensitivity to the $R$ and 
$\Delta xF_3$ at each ($x$, $Q^2$) bin comes from the $y$ dependence,
especially at high $y$. The values
of $F_2$ and $xF_3^{avg}$ at each $x$ and $Q^2$  bin
are extracted
by averaging over all the $y$ bins.  In this analysis the
extracted values of $R$ are highly 
correlated with the values $\Delta xF_3$ at low $x$.
However, since the dimuon results~\cite{dimu} 
provides an independent determination of the size of the
strange sea ($s$),
it is used to constrain the $\Delta xF_3$ in the overall fit.
We use
$\Delta xF_3=x\int^1_x C_3(\frac xy,Q^2)4[s(y,Q^2)-c(y,Q^2)]dy$,
where $C_3$ is the hard scattering coefficient. Here, 
the strange sea ($s$) from our dimuon data, and the
 charm sea ($c$)  from the  CTEQ2M~\cite{CTEQ} parametrization
($c$ is an order of magnitude smaller 
than $s$ in our kinematic region). 
These estimated values of $\Delta xF_3$ are used
as input in the extraction
of $R$, $F_2$, and $xF_3^{avg}$ (3-parameter fit).
To obtain a measurement of the strange sea which is independent
of the dimuon data, a 4-parameter fit is performed in which
the relative size of strange sea,
$\kappa (\equiv \frac{2s}{u_s+d_s})$ is allowed to vary. Systematic errors
are determined for each of the fit parameters by repeating
the fits and varying
the experimental and model parameters within their respective
uncertainties.
The following systematic uncertainties are investigated: 
the muon and hadron energy scale uncertainty($1\%$), the uncertainties
in the flux extraction, normalization (here only $F_2$ and $xF_3^{avg}$ 
are affected), and physics model parameters
(e.g. strange sea, and charm sea).

\begin{figure}
\centerline{\psfig{figure=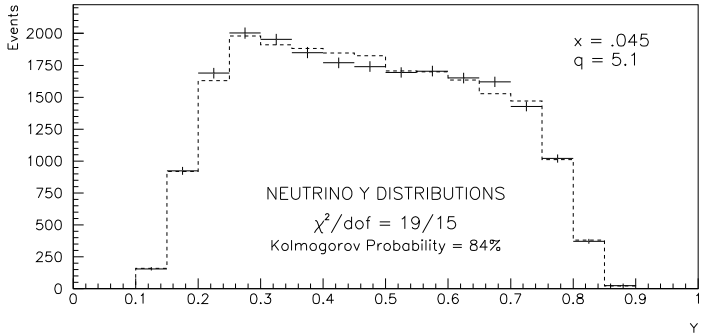,width=4.0in}}
\caption{The $y$-distribution for the $x=.045$, $Q^2=5.1$ GeV$^2$ bin.
The histogram is the data and the points 
\leftline{are the MC predictions
using the best fit structure functions 
from the 3-parameter fit
in that ($x,Q^2$) bin.}}
\label{fig:ydist}
\end{figure}
\begin{figure}
\centerline{\psfig{figure=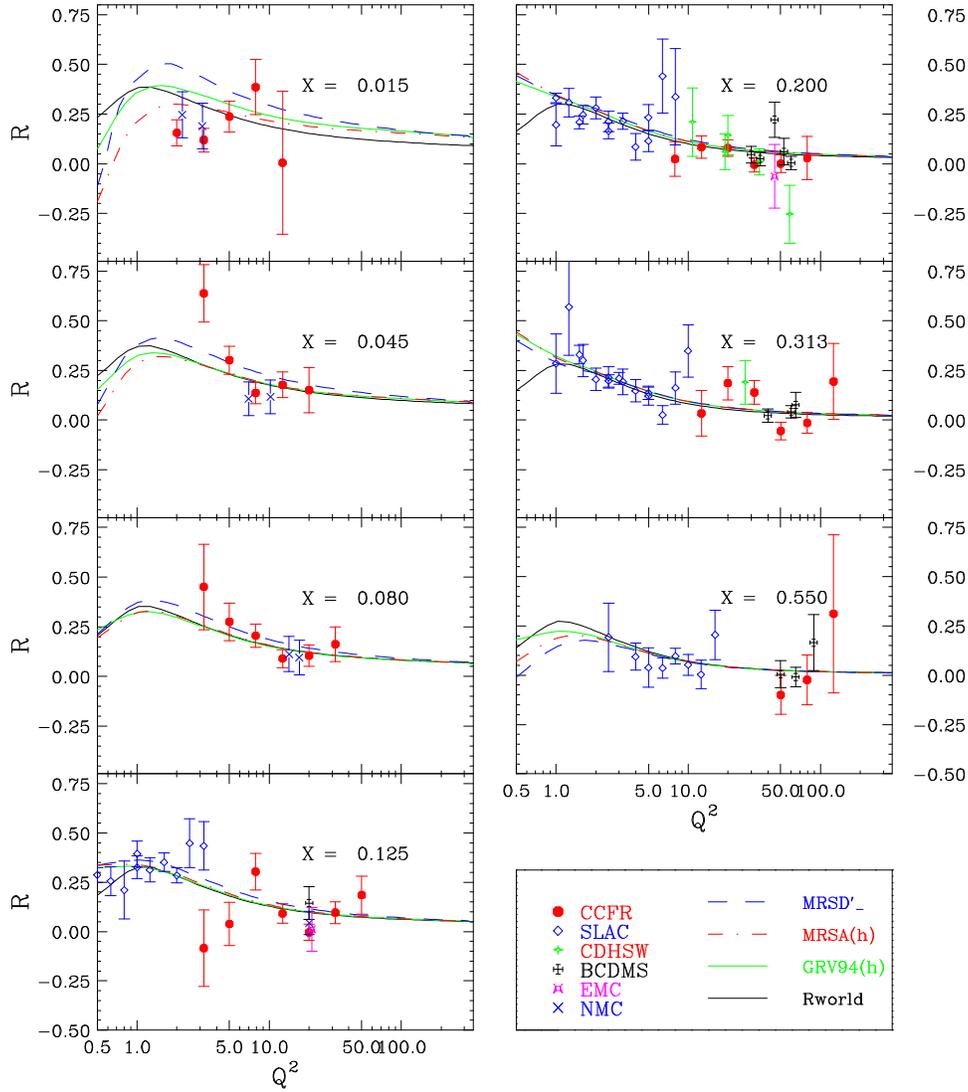,width=5.0in}}
\caption{Measurements of
 $R$ at fixed $x$ vs $Q^2$: The preliminary CCFR data are compared
with other measurements and
with a calculations from a QCD based model (Bodek, Rock and Yang)
using various parton 
\leftline{distribution functions.
The empirical Whitlow parameterization
(Rworld) is also shown as
a solid curve.}}
\label{fig:R}
\end{figure}

A typical  3-parameter fit to the $y$ distribution in one
of the ($x$ $Q^2$) bins 
is shown in Fig.~\ref{fig:ydist}.
The extracted values of $R$ at fixed $x$ vs $Q^2$ (from 3-parameter fit)
are shown on Fig.~\ref{fig:R}. The new data provide the first
$Q^2$ dependent measurements at $x<0.05$. The NMC data~\cite{rnmc} shown 
in Fig.~\ref{fig:R}
are integrated over $Q^2$, and plotted on the nearest $x$ bin.
At higher $x$,  our new measurements
are in good agreement with the other existing data 
(SLAC~\cite{rslac}, EMC~\cite{remc}, BCDMS~\cite{rbcdms}, NMC, 
and CDHSW~\cite{rcdhsw}),   
and with predictions~\cite{BRY} of
the Bodek, Rock and Yang calculated using various PDF's~\cite{PDF} (the predictions
come from
a QCD based model which includes NNLO terms,
heavy quark effects, target mass corrections, and a phenomenologically
determined higher twist contribution).
The extracted values of $F_2$ and $xF_3^{avg}$ from the 3-parameter fit
are in good agreement with previous CCFR measurements~\cite{SEL}
using these data.

\begin{figure}
\centerline{\psfig{figure=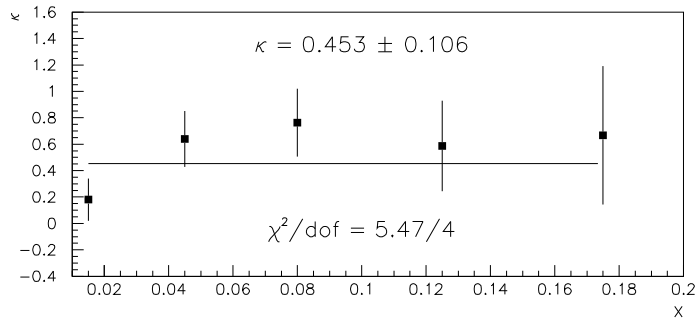,width=4.0in}}
\caption{Values of $\kappa$ extracted
from 4 parameter fit. Here $\kappa(x)$ is assumed
to be constant over the $Q^2$ range 
\leftline{in each $x$ bin. The solid line shows the 
average value of 
$\kappa=0.453 \pm 0.106_{-0.096}^{+0.028}$.}}
\label{fig:kappa}
\end{figure}

The  extracted values
of $\kappa$ from the 4-parameter
fit are shown in Fig.~\ref{fig:kappa} as a function of $x$.
 The values 
of $\kappa$ do not show any dependence on $x$.
The overall average value of $\kappa
=0.453 \pm0.106_{-0.096}^{+0.028}$ is in good agreement with the dimuon 
result ($\kappa = 0.477 \pm0.055$), indicating that the QCD effects
of strange sea in both inclusive and dimuon cross section
are consistent, thus the 3-parameter fit
provide a reliable extraction of $R$.
This measurement of $\kappa$ can be combined with our independent 
result~\cite{dimu}
of $\frac{\kappa}{\kappa+2}|V_{cs}|^2 =0.200\pm.015$ derived from the dimuon
analysis to obtain the CKM matrix element $|V_{cs}|$. The resulting
value of $|V_{cs}|$ is $1.05\pm0.10_{-0.11}^{+0.07}$.
This determination is relatively free of theoretical assumptions.
There is good agreement with the value of $|V_{cs}|_D=1.01\pm 0.18$ 
\cite{PHY} extracted from the $D_{e3}$ decay rate (that value relies 
on theoretical assumptions about the $D$ form factor at $Q^2=0$).

In summary, the new CCFR measurements of
$R$ extend to lower $x$ and higher $Q^2$
than previous results.
The extracted value of $\kappa$
is in good agreement with the value previously measured using
the dimuon data sample,
and yields the most precise experimental determination of 
$|V_{cs}|$. New data from our recent NuTeV run (96-97),
taken with a sign selected neutrino beam,
are expected to yield more precise determination of $R$
and $k$ at low $x$.

\end{document}